\documentclass[12pt]{article}
\usepackage[colorlinks=true,
            urlcolor=red,
            citecolor=blue]{hyperref}
\usepackage{amsmath}
\usepackage{enumerate}
\usepackage{natbib}
\usepackage[dvipdfmx]{graphicx}
\usepackage{fancybox,ascmac}
\usepackage{amsmath,amssymb,euscript,lscape,theorem,dcolumn,tascmac}
\usepackage{array,multirow,graphicx}
\newcommand{\blind}{0}
\begin{document}

\def\spacingset#1{\renewcommand{\baselinestretch}%
{#1}\small\normalsize} \spacingset{1}


\if0\blind
{
  \title{\bf Parallel Resampling \\ for Fully Parallelized Particle Filters}
  \author{Kenichiro McAlinn\thanks{Kenichiro McAlinn is Graduate Student, Department of Statistical Science, Duke University, Durham, NC 27708. (E-mail: \url{kenichiro.mcalinn@stat.duke.edu})} \hspace{.2cm}
\\
    Department of Statistical Science, Duke University\\
    and \\
    Teruo Nakatsuma \thanks{Teruo Nakatsuma is Professor, Faculty of Economics, Keio University, 2-15-45 Mita, Minato-ku, Tokyo, 108-8345, Japan (E-mail: \url{nakatuma@econ.keio.ac.jp}).}\\
    Faculty of Economics, Keio University}
  \maketitle
} \fi

\bigskip
\begin{abstract}
We develop a novel parallel resampling algorithm for fully parallelized particle filters, which is designed with GPUs (graphics processing units) or similar parallel computing devices in mind. With our new algorithm, a full cycle of particle filtering (computing the value of the likelihood for each particle, constructing the cumulative distribution function (CDF) for resampling, resampling the particles with the CDF, and propagating new particles for the next cycle) can be executed in a massively and completely parallel manner.  One of the advantages of our algorithm is that every single numerical computation or memory access related to the particle filtering is executed solely inside the GPU in parallel, and no data transfer between the GPU's device memory and the CPU's host memory occurs unless for further processing, so that it can circumvent the limited memory bandwidth between the GPU and the CPU. To demonstrate the advantage of our parallel algorithm, we conducted a Monte Carlo experiment in which we apply the parallel algorithm as well as conventional sequential algorithms for estimation of a simple state space model via particle learning, and compare them in terms of execution time. The results show that the parallel algorithm is far superior to the sequential algorithm.\\[1ex]
\end{abstract}

\noindent%

\spacingset{1.45}
\section{INTRODUCTION\label{intro}}

The state space model (SSM) has been one of the indispensable tools for time series analysis 
and optimal control for decades. Although the archetypal SSM is linear and Gaussian, 
the literature of more general non-linear and non-Gaussian SSMs has been 
rapidly growing in the last two decades. 
For lack of an analytically tractable way to estimate the general SSM, 
numerous approximation methods have been proposed. 
Among them, arguably the most widely applied method is 
particle filtering \citep{GSS93, Kitagawa96}. 
Particle filtering is a type of sequential Monte Carlo method in which the integrals we need to evaluate for filtering are approximated by the Monte Carlo integration.
To improve numerical accuracy and stability of the particle filtering algorithm, 
various extensions, such as the auxiliary particle filter \citep{PS99}, 
have been proposed, and still actively studied by many researchers. 
For SSMs with unknown parameters, 
\citet{Kitagawa98} proposed a self-organizing state space modeling approach in which 
the unknown parameters are regarded as a subset of the state variables and 
the joint posterior distribution of the parameters and the state variables is evaluated 
with a particle filtering algorithm. Other particle filtering methods that can simultaneously 
estimate parameters have been proposed by \citet{LW01}, \citet{Storvik02}, 
\citet{Fearnhead02}, \citet{PSM08}, \citet{JP08}, \citet{JPY08}, 
\citet{CJLP10}, just to name a few. These particle filtering methods that estimate state variables and parameters simultaneously are often called 
particle learning methods in the literature. 
Although the effectiveness of particle filtering methods have been proven through many different applications (see \citet{MTKW03}, \citet{ZC07}, \citet{MAHBCR07}, \citet{CY07}, \citet{LT11}, and \citet{DLP12} among others), 
it is offset by the fact that it is a time-consuming technique. 
Some practitioners still shy away from using it in their applications because of this despite its benefit.

This attitude toward particle filtering would be changed by the latest technology: 
Parallel computing. As we will discuss in Section 2, 
some parts of the particle filtering procedure are ready to be executed simultaneously 
on many processors in a parallel computing environment. 
In light of inexpensive parallel processing devices such as GPGPUs\footnote{A high-performance GPU  
(graphics processing unit) was originally developed for displaying 
high-resolution 2D/3D graphics necessary in video games and computer-aided design.
Because a GPU is designed with a massive number of processor cores 
to conduct single-instruction multiple-data (SIMD) processing, 
it has been regarded as an attractive platform of parallel computing and 
researchers started to use it for high-performance computing. 
As GPU manufacturers try to take advantage of this opportunity, 
it has evolved into a more computation-oriented device called GPGPU. Nowadays almost all GPUs have 
more or less capabilities for parallel computing, 
so the distinction between GPUs and GPGPUs are blurred.} 
(general purpose graphics processing units) available to the general public, 
more and more researchers start to jump on the bandwagon of parallel computing. 
\citet{SWCFCW10} and \citet{LYGDH10} reviewed general attempts at parallelization of Bayesian estimation methods. 
\citet{DG11} implemented a sequential Monte Carlo method on the GPU 
and applied it to complex nonlinear dynamic models, 
which are numerically intractable even for the Markov chain Monte Carlo method.

As for parallelization of particle filtering, a few researches 
(see \citet{MPSF04}, \citet{BDH05}, \citet{MAM06}, \citet{HHKG07}, \citet{HKG10}, \citet{CCCW10}, \citet{GBO12}, for example) 
have been reported, though the field is still in a very early stage. 
However, all of these state of the art methods are either 1) simple implementations onto parallel devices, 2) modifies the algorithm in a way that introduces additional estimation errors, or 3) depends on device-specific functionalities that would make it inapplicable for other devices. For example, \citet{LYGDH10} and \citet{DG11} are both parallel implementations of the particle filter, however, the resampling step for both implementations are computed sequentially and thus they are not fully parallel algorithms. \citet{HKG10} and \citet{GBO12}, on the other hand, use device-specific functionalities of the GPU to parallelize the resampling step that cannot be implemented in other devices. One method that has been used is what is called local resampling \citep{CCCW10}, which breaks up the resampling step in to several blocks and sequentially resample within that block. This method is obviously not a fully parallel algorithm as, while the computationally burden is lessened, it requires sequential computation within blocks, and thus not exploiting the full power of the parallel framework.
A fully parallel algorithm for particle filters have yet to be developed, to the best of the authors' knowledge.


In developing parallel algorithms, with particularly GPUs in mind, there are a few bottlenecks 
one should avoid. First, processing sequential algorithms on the GPU can be inefficient 
because of the GPU's device memory architecture and its lack of clock speed compared to the CPU. Roughly speaking, a GPU has two types of memory:
memory assigned to each core and memory shared by all cores. 
Access to the core-linked memory is fast while access to the shared memory 
takes more time. Generally, one should try as much to keep all calculations on each core 
without any large-scale communications among cores. The second bottleneck is that 
it is time-consuming to transfer memory between the host memory, which the CPU uses, and 
the device memory, which the GPU uses. 
In other words, the bandwidth between the GPU's device memory and the CPU's host memory 
is very narrow. We can see that a fully parallel algorithm defined above would be ideal for GPU devices as it would, automatically and without manipulation, be able to calculate everything 
within the GPU and without bottlenecks.

With these bottlenecks in mind, we have developed a new parallel resampling algorithm to complete the first fully parallel algorithm that computes 
the full cycle of the particle filtering algorithm in a massively and fully parallel manner.  
This includes the computing of the likelihood for each particle, 
constructing the cumulative distribution function (CDF) for resampling, 
resampling the particles with the CDF, and propagating new particles for the next cycle.
By keeping all of our computations within the GPU and 
avoiding all memory transfer between the GPU and the CPU 
during the execution of the particle filtering algorithm, we exploit the great benefits of parallel computing to the fullest while avoiding its short comings, especially on the GPU. 
While we designed our algorithm with GPUs in mind, since our parallel algorithm is a fully parallel algorithm,
it can be easily implemented on other parallel computing devices. 

In order to compare our new parallel algorithm with conventional sequential algorithms, 
we conduct a Monte Carlo experiment in which we apply the competing particle learning 
algorithms to estimate a simple state space model (stochastic trend with noise model)   
and record the execution time of each algorithm. The results show that our parallel algorithm 
on the GPU is far superior to the conventional sequential algorithm on the CPU by around 30$\times$. Focusing only on the resampling step, we have achieved a speed up of around 10$\times$, which considering the sequential nature of the algorithm, is a significant improvement.

The organization of this paper is as follows. In Section 2, we briefly review state space models 
and particle filtering and learning. In Section 3, we describe how to implement 
a fully parallelized particle filtering algorithm, in particular how to parallelize
the resampling step. In Section 4, we report the results of our Monte Carlo experiment and discuss their implications. 
In Section 5, we state our conclusion. 

\section{STATE SPACE MODELS AND PARTICLE FILTERING}

A general form of SSM is given by
\begin{equation}
\label{ssm}
\begin{cases}
y_t \sim p(y_t \vert x_t) \\
x_t \sim p(x_t \vert x_{t-1})
\end{cases}
\end{equation}
where $p(y_t|x_t)$ stands for the conditional distribution or density of observation $y_t$ 
given unobservable $x_t$ and 
$p(x_t|x_{t-1})$ stands for the conditional distribution or density of $x_t$
given $x_{t-1}$, which is the previous realization of $x_t$ itself. 
In the literature of SSM, unobservable $x_t$, which dictates the stochastic process of $y_t$, 
is called the state variable. 

Time series data analysis with the SSM is centered on how to dig 
up hidden structures of the state variable out of the observations $\{y_t\}_{t=1}^T$. 
In particular, the key questions in applications of SSM are 
(i) how to estimate the current unobservable $x_t$, 
(ii) how to predict the future state variables, and 
(iii) how to infer about the past state variables with the data currently available. 
These aspects of state space modeling are called
\emph{filtering}, \emph{prediction}, and \emph{smoothing}, respectively.

The filtering procedure, which is the main concern in our study, 
is given by the sequential Bayes filter:
\begin{align}
\label{one.step.ahead.prediction}
p(x_t \vert y_{1:t-1})
&=\int p(x_t \vert x_{t-1})p(x_{t-1} \vert y_{1:t-1})dx_{t-1}, \\
\label{filtering}
p(x_t \vert y_{1:t})
&=\frac{p(y_t \vert x_t)p(x_t \vert y_{1:t-1})}
{\displaystyle \int p(y_t \vert x_t)p(x_t \vert y_{1:t-1})dx_t},
\end{align}
where $y_{1:t}=\{y_1,\dots,y_t\}$ $(t=1,\dots,T)$ and  
$p(x_t|y_{1:t})$ is the conditional density of the state variable $x_t$ given $y_{1:t}$.
In essence, equation \eqref{filtering} is the well-known Bayes rule to update 
the conditional density of $x_t$ while equation \eqref{one.step.ahead.prediction} 
is the one-period-ahead predictive density of $x_t$ given the past observations $y_{1:t-1}$.
By applying \eqref{one.step.ahead.prediction} and \eqref{filtering} repeatedly, 
one keeps the conditional density $p(x_t \vert y_{1:t})$ updated as a new observation comes in.

In general, a closed-form of neither \eqref{one.step.ahead.prediction} nor \eqref{filtering} 
is available, except for the linear Gaussian case or when $x_t$ is finite where we can use the Kalman filter \citep{Kalman60}. 
See \citet{WH97} on detailed accounts about the linear Gaussian SSM. 
To deal with this difficulty, we apply particle filtering, 
in which we approximate the integrals in \eqref{one.step.ahead.prediction} and \eqref{filtering} 
with \emph{particles}, a random sample of the state variables 
generated from either the conditional density $p(x_t|y_{1:t})$ 
or the predictive density $p(x_t|y_{1:t-1})$.
Let $\{ x_{t}^{(i)} \}^{N}_{i=1}$ denote $N$ particles generated from $p(x_{t} \vert y_{1:t-1})$. 
We can approximate $p(x_t \vert y_{1:t-1})$ by 
\begin{equation}
p(x_{t} \vert y_{1:t-1}) \simeq \frac{1}{N}\sum ^{N}_{i=1}\delta(x_t-x_{t}^{(i)})
\end{equation}
where $\delta(\cdot)$ is the Dirac delta.
Then the filtering equation \eqref{filtering} is approximated by a discrete distribution
\begin{align}
\label{pf.resampling}
p(x_t \vert y_{1:t})
&\simeq \frac{p(y_t\vert x_t)\displaystyle \frac{1}{N}\sum^{N}_{i=1}
\delta(x_t-x_{t}^{(i)})}{\displaystyle \int p(y_t\vert x_t)\frac{1}{N} 
\displaystyle \sum^{N}_{i=1}\delta(x_t-x_t^{(i)})dx_t} 
= \sum_{i=1}^{N}w^{(i)}_{t}\delta(x_t-x_t^{(i)}), \\
w^{(i)}_t &= \displaystyle\frac{p(y_t\vert x_t^{(i)})}
{\sum^{N}_{i=1}p(y_{t}\vert x_t^{(i)})}. \nonumber
\end{align}
\eqref{pf.resampling} implies that the conditional density $p(x_t|y_{1:t})$ is 
discretized on particles $\{x_t^{(i)}\}_{i=1}^N$ with probabilities $\{w_t^{(i)}\}_{i=1}^N$. 
Therefore we can obtain a sample of $x_t$, $\{\tilde x_{t}^{(i)} \}^{N}_{i=1}$, 
from $p(x_t|y_{1:t})$ by drawing each $\tilde x_{t}^{(i)}$ out of $\{x_t^{(i)}\}_{i=1}^N$ 
with probabilities $\{w_t^{(i)}\}_{i=1}^N$ 
when the approximation \eqref{pf.resampling} is sufficiently accurate. 
This procedure is called \emph{resampling}. 
In reverse, if we have $N$ particles $\{\tilde x_{t-1}^{(i)} \}^{N}_{i=1}$ generated from 
$p(x_{t-1} \vert y_{1:t-1})$, we can approximate $p(x_{t} \vert y_{1:t-1})$ by 
\begin{equation}
\label{pf.propagation}
p(x_{t}|y_{1:t-1})\simeq \int p(x_{t}\vert x_{t-1})
\frac{1}{N} \displaystyle \sum^{N}_{i=1}\delta(x_{t-1}-\tilde x_{t-1}^{(i)})dx_{t-1}
= \frac1N\sum_{i=1}^N p(x_{t}|\tilde x_{t-1}^{(i)}).
\end{equation}
Then \eqref{pf.propagation} implies that we can obtain a sample of $x_{t+1}$, 
$\{x_{t+1}^{(i)} \}^{N}_{i=1}$, 
from $p(x_{t+1}|y_{1:t})$ by generating each $x_{t+1}^{(i)}$ from $p(x_{t+1}|\tilde x_t^{(i)})$, 
which is called \emph{propagation}. 
Hence we can mimic the sequential Bayes filter by repeating the \emph{propagation} equation
\eqref{pf.propagation} and the \emph{resampling} equation \eqref{pf.resampling} 
for $t=1,2,3,\dots$  
This is the basic principle of particle filtering. 
The formal representation of the particle filtering algorithm is given as follows.
\newline
\newline
\textsc{Algorithm: Particle Filtering}
\begin{description}
\item[\hspace{0.75cm}Step 0:] Sample the starting values of $N$ particles $\{ \tilde x_{0}^{(i)} \}^{N}_{i=1}$ from $p(x_0)$.
\item[\hspace{0.75cm}Step 1:] Propagate $x_{t}^{(i)}$ from  $p(x_{t} \vert \tilde{x}_{t-1}^{(i)}),$ 
	$(i=1,\dots ,N)$.
\item[\hspace{0.75cm}Step 2:] Compute weight $w_{t}^{(i)} \propto p(y_{t} \vert x_{t}^{(i)})$ such that $\sum^{N}_{i=1}w_t^{(i)}=1$.
\item[\hspace{0.75cm}Step 3:] Resample $\{\tilde{x}_{t}^{(i)} \}^{N}_{i=1}$ from 
	$\{ x_{t}^{(i)} \}^{N}_{i=1}$ with weight $w_{t}^{(i)}$.
\end{description} 
\hspace*{\fill} 

When a state space model depends on unknown but fixed parameters $\theta$, 
\begin{equation}
\begin{cases}
y_t \sim p(y_t \vert x_t,\theta) \\
x_t \sim p(x_t \vert x_{t-1},\theta)
\end{cases}
\end{equation}
we need to evaluate the posterior distribution $p(\theta|y_{1:t})$ 
given the observations $y_{1:t}$. In the framework of particle filtering, $p(\theta|y_{1:t})$ is sequentially updated as a new observation arrives, which is called particle learning.
The particle learning algorithm is defined as follows.
Let $\{z_{t}^{(i)}=(x_{t}^{(i)},\theta_{t}^{(i)})\}_{i=1}^N$ 
and $\{\tilde z_t^{(i)}=(\tilde x_t^{(i)},\tilde \theta_t^{(i)})\}_{i=1}^N$ 
denote particles jointly generated from $p(x_{t},\theta|y_{1:t-1})$ and 
$p(x_{t},\theta|y_{1:t})$ respectively.
Then the particle approximation of the Bayesian learning process (Kitagawa (1998)) is given by
\begin{align}
p(z_{t}|y_{1:t-1})&\simeq \frac1N\sum_{i=1}^N p(z_{t}|\tilde z_{t-1}^{(i)}),\\
p(z_t\vert y_{1:t})
&\simeq \sum_{i=1}^{N}w^{(i)}_{t}\delta(z_{t}-z_{t}^{(i)}), \quad
w^{(i)}_{t} = \displaystyle 
\frac{p(y_{t}\vert z_{t}^{(i)})} {\sum^{N}_{i=1}p(y_{t}\vert z_t^{(i)})}.
\end{align}
This is a rather straightforward generalization of particle filtering. We must note that since $\theta$ appears in both sides of the distribution $p(z_t|\tilde{z}_{t-1}^{(i)})$, we are implicitly treating the static parameter $\theta$ as a time-varying one and thus we cannot incorporate it as simply an augmented state. Some state-of-the-art particle methods such as PMCMC by \citet{ADH10} and  $SMC^2$ by \citet{CJP13}, for example, address this problem.

The particle learning algorithm by \citet{CJLP10} learns the parameters by deterministically updating the sufficient statistics of the parameter distribution, denoted by $s_t$, via the recursion map, $\mathcal{S}(\cdot)$. The formal expression for the particle learning algorithm is summarized as follows.
\newline
\newline
\textsc{Algorithm: Particle Learning}
\begin{description}
\item[\hspace{0.75cm}Step 0:] Sample the starting values of $N$ particles $\{ z_{0}^{(i)} \}^{N}_{i=1}$ from $p(z_0)$.
\item[\hspace{0.75cm}Step 1:] Resample $\{\tilde{z}^{(i)}_t\}^{N}_{i=1}$ from $z^{(i)}_t=(x_t,s_t,\theta)^{(i)}$ with weights $w_t\propto p(y_{t+1}|z^{(i)}_t)$ such that $\sum^{N}_{i=1}w_t^{(i)}=1$.
\item[\hspace{0.75cm}Step 2:] Propagate $\tilde{x}^{(i)}_{t}$ to $x^{(i)}_{t+1}$ via $p(x_{t+1}|\tilde{z}^{(i)}_t,y_{t+1})$.
\item[\hspace{0.75cm}Step 3:] Update sufficient statistics $s^{(i)}_{t+1}=\mathcal{S}(\tilde{s}^{(i)}_{t},x^{(i)}_{t+1},y_{t+1})$.
\item[\hspace{0.75cm}Step 4.] Sample $\theta^{(i)}$ from $p(\theta|s^{(i)}_{t+1})$.
\end{description} 
\hspace*{\fill} 

Once we generate  $\{\tilde\theta_t^{(i)}\}$ by the particle learning algorithm, 
we can treat them as a Monte Carlo sample of $\theta$ 
drawn from the posterior density $p(\theta|y_{1:t})$. 
Thus we calculate the posterior statistics on $\theta$ with $\{\tilde\theta_t^{(i)}\}$ 
in the same manner as the traditional Monte Carlo method.

The computational burden of particle filtering will be prohibitively taxing 
as the number of particles $N$ increases. 
The number of likelihood evaluations,
the number of executions for constructing the CDF of particles, 
and the number of particles to be generated in propagation is $\mathcal{O}(N)$.
The number of executions for resampling with the CDF will increase in $\mathcal{O}(N^2)$
when we use a naive resampling algorithm, but it can be reduced to $\mathcal{O}(N\log N)$ 
with more efficient algorithms. 
We can see that sequential particle-by-particle execution of each step in the particle filtering (and learning) algorithm
is inefficient when $N$ is large, while the particle filtering method by construction 
requires a large number of particles to guarantee precision in the estimation.

To reduce the time for computation,  
we propose to parallelize the resampling step in order to parallelize all steps in particle filtering so that
we can execute the parallelized particle filtering algorithm completely inside the GPU. 
The key to constructing an efficient parallel algorithm is asynchronous out-of-order execution 
of jobs assigned to each processor.
We need to keep a massive number of processors in the GPU as busy as possible to fully exploit 
the potential computational power of the GPU. Therefore, each processor should waste no milliseconds 
by waiting for other processors to complete their jobs. 
If the order of execution is independent of the end result, 
asynchronous out-of-order execution is readily implemented.
In this situation, parallelization is rather straightforward. 
In the particle filtering method, this is the case for computing the likelihood and 
the propagation step and these steps can be computed in parallel without any modifications (i.e., parallel in nature). 
For constructing the CDF and resampling particles, on the other hand, 
the conventional algorithm does not allow asynchronous out-of-order execution and 
thus parallelization is difficult to achieve. 
In order to devise a fully parallelized particle filtering method, 
these steps must be fully parallelized as well.
The parallelization of the CDF construction, or the parallelization of the cumulative sum, has been investigated extenssively (see for example, \citet{HHKG07}). In our parallel particle filter algorithm, we utilize the parallel construction of the CDF by \citet{HHKG07}, called the forward adder backward adder algorithm, and develop a new fully parallel algorithm for resampling. While the parallel CDF construction by \citet{HHKG07} may not be the most efficient parallel algorithm, the examination and comparison of these algorithms are beyond the scope of this paper as it has little to do with our innovation with our parallel resampling algorithm.
In the next section, we describe how to implement resampling 
in a parallel computing environment. 

\section{FULL PARALLELIZATION OF PARTICLE FILTERING}

\subsection{A Review of the Conventional Resampling Algorithms}

The goal of resampling is to generate $N$ random integers, which are the indices of particles,  
from a discrete distribution on $\{1,\ldots,N\}$ with the cumulative distribution function,
\begin{equation}
\label{resample.cdf}
q(i) \propto \sum_{j=1}^i w_t^{(j)}, \quad (i = 1,\dots, N).
\end{equation}
Many resampling methods have been invented through the years and its benefits and drawbacks have been thoroughly examined (see \citet{DCM05}, for example). Rather than going over the details of such resampling methods, we will focus our review in regards of parallel computing before we proceed to describe our parallel resampling algorithm. In particular, we will review three popular sampling methods, multinomial, stratified, and systematic resampling.
\newline
\newline
\textsc{Multinomial Resampling}

The multinomial resampling is the most basic and exact resampling procedure, however, 
it is an extremely time-consuming $\mathcal{O}(N^2)$ operation, 
as we need to search through the whole PDF for each particle sequentially.
Yet this resampling method would be ideal for parallel contexts as each sample from the uniform distribution can be drawn in parallel. Additionally, it is the only resampling method with theoretical justification (see \citet{Chopin04} for the central limit theorem proof for the particle filter with multinomial resampling) and thus ideal and exact if we set the number of particles to be sufficiently large.

In sequential computing, the extremely time-consuming nature of the multinomial resampling can be averted by sorting the random samples from the uniform distribution in ascending order. 
By sorting the uniform variates, each search for a particle 
can be started where the last research left off. The offset of this algorithm is that, 
although the resampling procedure is a more efficient $\mathcal{O}(N\log N)$ operation, 
sorting uniform variates can be computationally strenuous as the number of particles increases, depending on the sorting algorithm we use. Additionally, once sorted, there are little benefits in parallelization.
\newline
\newline
\textsc{Stratified Resampling}

The stratified resampling conducts the resampling procedure by generating uniform variates 
on $N$ equally spaced intervals $[(i-1)/N,i/N]$ $(i=1,\dots,N)$. 
Since only one particle will always be picked for each interval, it is a pseudo-random variate and
it does not exactly generate random integers from the the CDF $\{q(1),\dots,q(N)\}$. Although it benefits from lesser Monte Carlo error, it lacks the theoretical justification and is unsuitable for parallelization as its benefit come from its sequential nature.
\newline
\newline
\textsc{Systematic Resampling}

The systematic resampling is similar to the stratified resampling 
but it always chooses an identical point in $[(i-1)/N,i/N]$ for all $i$. This method suffers from the same problems with stratified resampling; it is theoretically unjustified as it only produces one random sample for all samples. It is especially problematic when the distribution of the parameter is complex and many of the points in the CDF falls between $[(i-1)/N,i/N]$.

When we consider parallelization of the resampling algorithm, the multinomial resampling is the best suited as the generation of random variates is done in parallel and is the only theoretically justified method, in the sense of asymptotics pointed out in \citet{Chopin04} (i.e., the lack of asymptotic theory for the later two resampling techniques). The benefits of using the other two resampling algorithms is only under sequential machines and when we want to lessen the number of particles without increasing Monte Carlo error. However, under the parallel paradigm, we can run the particle filter with massive number of particles under a very reasonable amount of time, thus there is no reason to use the stratified and systematic resampling in parallel contexts. However, even though the benefits are small, our parallel resampling algorithm can be used for stratified and systematic resampling as well.

\subsection{Fully parallelized resampling}
Previous parallel resampling algorithms for particle filtering were either specifically designed for GPUs (\citet{HKG10}, \citet{GBO12}) utilizing its specific functionalities or were not fully parallel (\citet{CCCW10}). 
The method of \citet{HKG10}, for example, is dependent on a device specific functionality (rasterization) and 
its efficiency and scalability is limited to the GPU architecture. 
Our parallel algorithm, on the other hand, is more versatile and scalable because 
it requires only basic thread coordination mechanisms such as shared memory and 
thread synchronization which are provided by most parallel computing systems.
Our method to parallelize the resampling procedure while maintaining its exactness, 
we have developed a parallel resampling algorithm based on the cut-point method by \citet{CA74}, which is our main contribution. 

A \emph{cut-point}, $I_j$, for given $j=1,\dots,N$ is the smallest index $i$ such that 
the corresponding probability $q(I_j)$ should be greater than $(j-1)/N$. 
In other words, 
\begin{equation}
I_j=min\left\{i:q(i)>\frac{j-1}{N}\right\}, \quad (j=1, \dots,N).
\end{equation}
Given cut-points $\{I_1,\dots,I_N\}$,
random integers between 1 and $N$ is generated from the CDF
by the following procedure:
\newline
\newline
\textsc{Algorithm: Cut-Point Method}
\begin{description}
\item[\hspace{0.75cm}Step 0:] Let $j=1$ 
\item[\hspace{0.75cm}Step 1:] Generate $u$ from the uniform distribution on the interval $(0,1)$.
\item[\hspace{0.75cm}Step 2:] Let $k=I_{\lceil Nu\rceil}$ where $\lceil x\rceil$ stands for 
	the smallest integer greater than or equal to $x$.
\item[\hspace{0.75cm}Step 3:] If $u>q(k)$, let $k\leftarrow k+1$ and repeat \textbf{Step 3}; 
	otherwise, go to \textbf{Step 4}. 
\item[\hspace{0.75cm}Step 4:] Store $k$ as the index of the particle.
\item[\hspace{0.75cm}Step 5:] If $j<N$, let $j\leftarrow j+1$ and go back to \textbf{Step 1}; 
	otherwise, exit the loop.
\end{description}
\hspace*{\fill} 

Once all cut-points $\{I_1,\dots,I_N\}$ are given, 
parallel execution of the cut-point method
is straightforward because the execution of \textbf{Step 1} -- \textbf{Step 3} 
does not depend on the index $j$. 
The fully parallel resampling algorithm 
distributed on $N$ threads is given as follows.
\newline
\newline
\textsc{Algorithm: Parallelized Cut-Point Method}
\begin{description}
\item[\hspace{0.75cm}Step 0:] Initiate the $j$-th thread.
\item[\hspace{0.75cm}Step 1:] Generate $u$ from the uniform distribution on the interval $(0,1)$.
\item[\hspace{0.75cm}Step 2:] Let $k=I_{\lceil Nu\rceil}$ where $\lceil x\rceil$ stands for 
	the smallest integer greater than or equal to $x$.
\item[\hspace{0.75cm}Step 3:] If $u>q(k)$, let $k\leftarrow k+1$ and repeat \textbf{Step 3}; 
	otherwise, go to \textbf{Step 4}. 
\item[\hspace{0.75cm}Step 4:] Store $k$ as the index of the particle.
\item[\hspace{0.75cm}Step 5:] Wait until all threads complete the job.
	Otherwise, exit the loop.
\end{description}
\hspace*{\fill} 

However, the conventional algorithm for computation of the cut-points 
(see \citet[p.158]{Fishman96} for example) is not friendly to parallel execution.
Thus, we have developed an efficient algorithm for parallel search of all cut-points.
To devise such a search algorithm, let us define
\[
 L_j = \lceil Nq(j)\rceil, \quad (j=1,\dots,N)
\]
and $L_0=0$. Due to the monotonicity of the CDF, we observe 
\begin{enumerate}
\item $0=L_0< L_1\le \cdots \le L_N=N$.
\item If $L_{j-1}<L_j$, a cut-point such that 
\begin{equation*}
I_k=\min\left\{i:q(i)>\frac{j-1}{N}\right\},
\quad
 (k=L_{j-1}+1,\dots,L_j)
\end{equation*}
is given as $I_k=j$. 
\item If $L_{j-1}=L_j$, $j$ is not corresponding to any cut-points.
\end{enumerate}
The above properties give us a convenient criterion to check whether a particular $L_j$ 
is a cut-point or not and it leads to 
the following multi-thread parallel algorithm to find all cut-points.
\newline
\newline
\textsc{Algorithm: Parallelized Cut-Point Search}
\begin{description}
\item[\hspace{0.75cm}Step 0:] Initiate the $j$-th thread.
\item[\hspace{0.75cm}Step 1:] Compute $L_j = \lceil Nq(j)\rceil$.
\item[\hspace{0.75cm}Step 2:] Let $k=L_j$.
\item[\hspace{0.75cm}Step 3:] If $k>L_{j-1}$, let $I_k=j$; otherwise, terminate the thread.
\item[\hspace{0.75cm}Step 4:] Let $k\leftarrow k-1$ and go to \textbf{Step 3}. 
\end{description}
\hspace*{\fill} 

To better illustrate our algorithm, we will give a simple example with $N=10$ in Table \ref{table.sample}. We suppose the CDF in Table \ref{table.sample} was already computed and stored in the GPU's device memory. We first initiate 10 threads and assign the $j$-th thread to compute $L_j = \lceil Nq(j)\rceil$,  $(j=1,\dots,10)$. Then the thread index $j$ is stored as a cut-point if $L_j>L_{j-1}$ holds, that is, $I_{L_j}=j$. In Table \ref{table.sample}, this is the case for $j=1,2,4,6,7,8,9,10$. For instance, since $L_2=3$ is greater than $L_{2-1}=L_1=2$, so we have $I_2=2$. Note that we have already found 8 cut-points $\{I_1,I_2,I_4,I_6,I_7,I_8,I_9,I_{10}\}$ by applying Step 3 in the cut-point search algorithm once. The remaining cut-points, $I_3$ and $I_5$, will be searched by repeating Step 3--4 in the algorithm. We have $I_{L_3-1}=I_2=2$ and $I_{L_5-1}=I_4=4$. In this example of the cut-point search, threads need to repeat Step 3--4 at most twice in order to find all cut-points.

We then assign each thread to generate a uniform random variate $u$, compute $I_{\lceil Nu\rceil}$, and search the index for resampling. For instance, the first thread picks the particle indexed by 1 because $u=0.0020<q(I_1)=q(1)=0.1182$. For the $9$-th thread, on the other hand, $u=0.1481$ is still greater than $q(1)=0.1182$, so it moves up the ladder of the CDF and finds $u<q(2)=0.2350$. Thus picking the particle indexed by 2. Note that
the number of iterations in each thread is at most two in this illustration of the parallel resampling procedure. 

Our parallel resampling algorithm will work most efficiently
if $(j-1)/N<q(j)\le j/N$ for all $j=1,\dots,N$. In this best-case scenario, both cut-point search and resampling are executed in one step for all threads since all cut-points are distinct. This means that all sampled uniform random variates will fall between distinct cut-points and thus there will be no need for searching through the CDF (i.e., climbing up the ladder). On the other hand, our cut-point search algorithm will be inefficient when the distribution of resampling weights are concentrated at a few points. Note that this occurs when particles degenerate. In this worst-case scenario, the CDF has a few but very high ``cliffs'' at which the thread is busy executing Step 3--4 in the cut-point search for many times, while other threads are idling because most of $L_j$'s are identical (thus the search ends instantaneously). In such case we can see that the benefit of parallelization is offset when the number of particles is extremely large compared to the number of distinct values among particles. Fortunately, this inefficiency is always associated with particle degeneracy, which we should and can avoid in practice by resampling at every step and periodically reinitiating the particle filter.

\begin{table}[htbp]
\caption{Parallel cut-point resampling with N=10}
\label{table.sample}
\begin{center}
\begin{tabular}{r|cccc|ccc|c}
\shortstack{Thread\\ Index}&\shortstack{CDF\\ $q(j)$}&$L_j$&$L_j>L_{j-1}$&$I_j$&$u$&$\lceil Nu\rceil$&$I_{\lceil Nu\rceil}$&\shortstack{Resampled\\ Index}\\
\hline
$j=$
1&0.1182&2&$\circ$&1&0.0020&1&1 &1\\
2&0.2350&3&$\circ$&2&0.2974&3&2 &4\\
3&0.2971&3&$\times$&2&0.0421&1&1 &1\\
4&0.4053&5&$\circ$&4&0.7461&8&8 &8\\
5&0.4571&5&$\times$&4&0.4011&5&4 &4\\
6&0.5109&6&$\circ$&6&0.5377&6&6 &7\\
7&0.6258&7&$\circ$&7&0.7145&8&8 &8\\
8&0.7583&8&$\circ$&8&0.6732&7&7 &8\\
9&0.8659&9&$\circ$&9&0.1481&2&1 &2\\
10&1&10&$\circ$&10&0.8691&9&9 &10\\
\end{tabular}
\end{center}
\end{table}

With a fully parallel resampling algorithm, particle filtering can be executed 
in a fully parallel manner without any compromise or inefficiency. Additionally, 
as the particle filtering (learning) algorithm is 
conducted completely on the GPU and each particle goes through the algorithm 
on each designated core without syncing, the advantage of parallel computing is gained 
to the fullest while its shortcoming is kept to its minimum 
(data transfer between the GPU's device memory and the CPU's host memory only occurs 
at the beginning and the end of the computation).

\begin{center}
\section{NUMERICAL EXPERIMENT}
\end{center}

In our experiment, we use a stochastic trend with noise model:
\begin{equation}
\label{trend+noise}
\begin{cases}
y_t = x_t + \nu_t, \quad \nu_t \sim {\cal N}(0,\sigma^2)\\
x_t = x_{t-1} + \epsilon_t, \quad \epsilon_t \sim {\cal N}(0,\tau^2)
\end{cases}
\end{equation}
as the benchmark model for performance comparison. 
In \eqref{trend+noise}, we set $x_0=0$, $\sigma^2=1$, $\tau^2=0.1$, and 
generate $\{y_1,\dots,y_{100}\}$. 
Then we treat $\sigma^2$ and $\tau^2$ as unknown parameters and 
apply the particle learning algorithm by \citet{CJLP10} to \eqref{trend+noise}.
The prior distributions are 
\begin{equation*}
 x_0\sim\mathcal{N}(0,10),\quad 
 \sigma^2\sim\mathcal{IG}(5,4),\quad
 \tau^2\sim\mathcal{IG}(5,0.4)
\end{equation*}

To demonstrate the effectiveness of our new parallel algorithm, 
we will compare the following types of algorithms:
\begin{itemize}
\item Sequential algorithm on the CPU
\begin{description}
\item[\emph{CPU(s)}:] Resampling with sorted uniform variates with single precision
\end{description}
\item Parallel algorithm on the GPU
\begin{description}
\item[\emph{GPU(sp)}:] Parallel resampling by the cut-point method with single precision
\item[\emph{GPU(dp)}:] Parallel resampling by the cut-point method with double precision
\end{description}
\end{itemize}
The first is a conventional sequential algorithm for resampling. 
The code for the parallel algorithm is written in CUDA while that for the sequential algorithms 
is in C. Both are compiled and executed on the same Linux PC\footnote{For our applications we use \texttt{CUDA} version 5.0 and compiled on \texttt{gcc} version 4.4.3}. For CPU computations, we have used an Intel Core-i7 2700k with 3.50GHz of clock rate with four cores and 16GB of memory. The GPU computation was done on a NVIDIA GTX580 with 772MHz of clock rate and 512 cores with 3GB of memory.
Alternative resampling algorithms, such as the stratified and systematic resampling, 
are not considered here, as they are not exact and do not benefit from parallel frameworks, as explained in Section 3.2. 
However, if we exclude the time consumed by the sorting procedure 
from the resampling time of \emph{CPU(s)}, 
we get a very good estimate of how long they might take. We also add that for sequential resampling methods, running the algorithm on the CPU will always outperform the GPU because of its core clock rate. We can thus think of the sequential resampling on the CPU after sorting to be the fastest alternative to the parallel resampling algorithm on the GPU (e.g. transferring the data from the GPU to CPU for the resampling step without the time it takes for memory transfer, which is extremely taxing for GPU computations).

\begin{figure}[htbp]
\centering\includegraphics[width=0.9\textwidth]{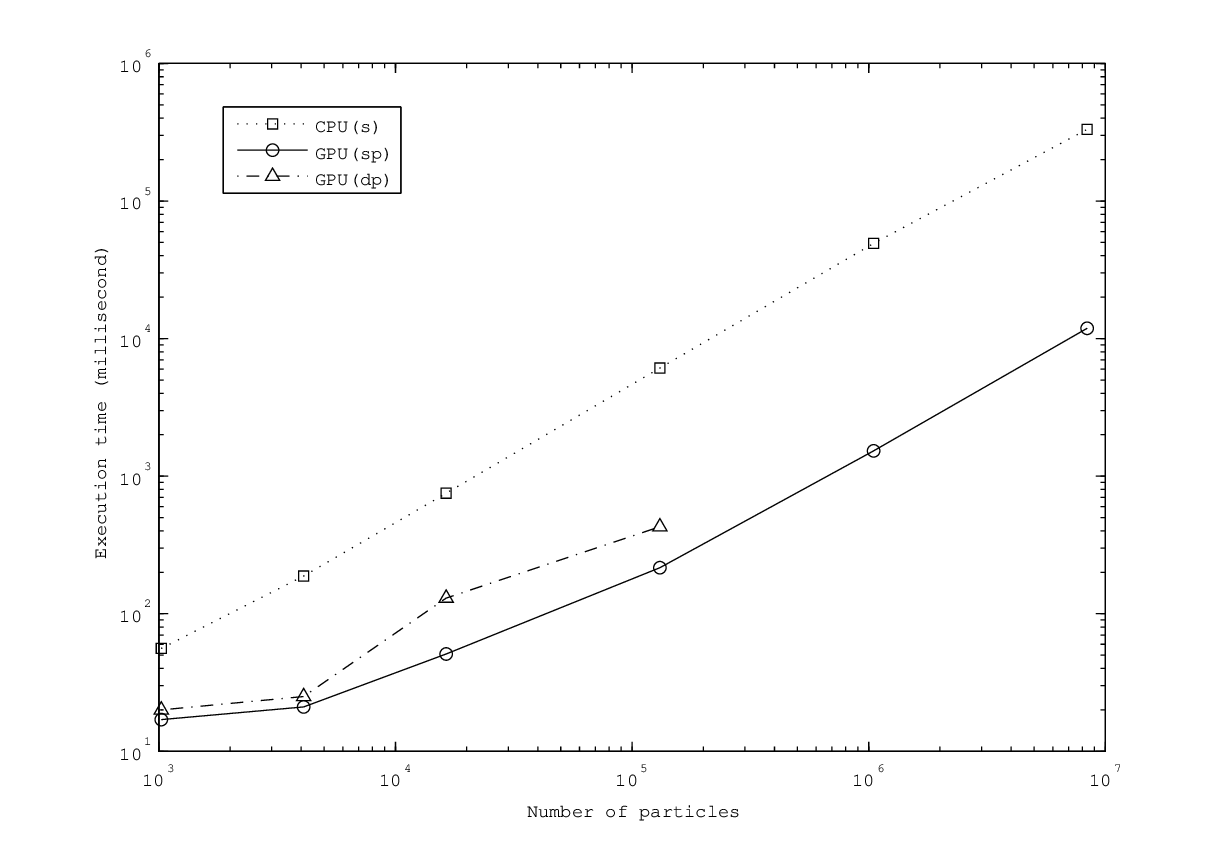}
\vspace{-0.5cm}
\caption{Plots of execution time against the number of particles}
\label{fig.comp}
\end{figure}

For each algorithm, we execute the particle learning ten times with  
the same generated path, $\{y_1,\dots,y_{100}\}$, and recorded the execution time of each trial.  
To avoid the influence of possible outliers, we took the average of the five closest to the median. 
The results are listed  in Table \ref{table.comp} and 
the plots of the total execution time against the number of particles are shown 
in Figure \ref{fig.comp}.

\begin{table}[htbp]
\caption{Comparison in execution time}
\label{table.comp}
\begin{center}
\begin{tabular}{ll|rrrrrrr}
\hline
\multicolumn{2}{r}{Number of particles:} & 1,024 & 4,096 & 16,384 & 131,072 & 1,048,576 & 8,388,608  \\ \hline
\parbox[t]{2mm}{\multirow{3}{*}{\rotatebox[origin=c]{90}{Time}}}&(i) \emph{CPU(s)} & 56 & 188 & 752 & 6,108 & 49,218 & 332,226 \\ 
&(ii) \emph{GPU(sp)} & 17 & 21 & 51 & 216 &	1,527 	&11,878   \\
&(iii) \emph{GPU(dp)} & 20 & 25 & 130 & 429 &	--- &	--- \\
\hline
\multicolumn{2}{r}{Number of particles:} & 1,024 & 4,096 & 16,384 & 131,072 & 1,048,576 & 8,388,608  \\ \hline
\parbox[t]{2mm}{\multirow{3}{*}{\rotatebox[origin=c]{90}{Ratio}}}&(i) / (ii) & 3.2  &	9.0  &	14.6 &	28.3 & 32.2 & 28.0  \\
&(i) / (iii) & 2.8 & 7.4 & 5.7 & 14.2  &--- &--- \\
&(iii) / (ii) & 1.1 & 1.2 & 2.5 & 2.0 &	---&---	 \\ \hline
\multicolumn{7}{l}{Note: The values of execution time are in milliseconds.}\\
\end{tabular}
\end{center}
\end{table}

The results clearly show that our new parallel algorithm, 
which runs completely in parallel and keeps all executions within the GPU, 
can be extremely effective and efficient compared to conventional sequential algorithms. 
As the number of  particles increases (and the precision of the estimate increases), 
\emph{GPU(sp)} consistently outperforms \emph{CPU(s)} by more than 20$\times$ when the number of particles is more than 131,072; 32$\times$ 
in the case of 1,048,576($=2^{20}$) particles and 
28$\times$ in the case of 8,388,608($=2^{23}$) particles. 
In the comparison between \emph{GPU(sp)} and \emph{GPU(dp)}, 
the difference is somewhere around two-fold, which is consistent with intuition. 
Interestingly, the computation on the GPU in double precision is still a good 5-10$\times$ faster than 
that of the CPU in single precision, 
which demonstrates the sheer power of parallel processing on the GPU. 
Due to memory failure, \emph{GPU(dp)} failed for particles more than a million, 
though this could be remedied by upgrading the GPU to the one with more memory 
or using multiple GPUs.

To see which part of the particle learning contributes to the reduction in execution time, 
we divide the cycle of particle learning into the following steps;
\begin{description}
\item[\hspace{0.75cm}\emph{Initialize}:] set the starting values of the particles;
\item[\hspace{0.75cm}\emph{CDF}:] compute the likelihood and construct the CDF;
\item[\hspace{0.75cm}\emph{Resample}:] resample the particles with the CDF;
\item[\hspace{0.75cm}\emph{Propagate}:] propagate a new set of particles;
\item[\hspace{0.75cm}\emph{Store}:] store the generated particles into the CPU's host memory (GPU only);
\item[\hspace{0.75cm}\emph{Other}:] keep the results and proceed with the particle learning;
\end{description}
The results in the case of 131,972 particles are listed in Table \ref{table.breakdown}. 
The tendency we observe in Table \ref{table.breakdown} is similar in the other cases.

\begin{table}[htbp]
\caption{Breakdown of execution time}
\label{table.breakdown}
\begin{center}
\begin{tabular}{l|r|rrrrr|r}
\cline{2-8}
 & & \multicolumn{5}{c|}{Cycle of particle learning} & \\
\hline
Time & Initialize & CDF & Resample & Propagate  & Store & Other & Total \\
\hline
(i) \emph{CPU(s)} & 26 & 2668 & 356 & 2912 & --- & 46 & 6108 \\
 & & & (92) & & & & \\
(ii) \emph{GPU(sp)} & 0.72 & 11 & 38 & 64 & 46 & 56 & 216 \\
(iii) \emph{GPU(dp)} & 1.51 & 19 & 54 & 174 & 89 & 92 & 429 \\
\hline
Ratio & Initialize & CDF & Resample & Propagate  & Store & Other & Total \\
\hline
(i) / (ii) & 36.1 & 242.5 & 9.4 & 45.5 & --- & 0.8 & 28.3 \\
 & & & (2.4) & & & & \\
(iii) / (ii) & 2.1 & 1.8 & 1.4 & 2.7 & 1.9 & 1.6 & 2.0 \\
\hline
\end{tabular}
\begin{tabular}{rl}
Notes: (a) &\!\!\!\!\!\! the number of particles is 131,072;\\
 (b) &\!\!\!\!\!\! the values of execution time are in milliseconds; \\
 (c) &\!\!\!\!\!\! the number in parentheses corresponds to the time excluding \\
 &\!\!\!\!\!\! the sorting step.
\end{tabular}
\end{center}
\end{table}

Breaking down the execution time gives us deep insights into how the GPU architecture works 
and its strong and weak points. Examining the results in \emph{CPU(s)}, 
we first notice that the \emph{CDF} step and \emph{Propagation} step put together occupy 
the bulk of the total execution time,  
while the \emph{Resampling} step only accounts for less than ten percent of 
the total execution time and much of it coming from the sorting step. 
Looking closely into the gains by parallelization in each step, 
the largest comes from the \emph{CDF} step with a gain of 248$\times$, 
followed by the \emph{Propagation} step with a gain of 45.3$\times$, 
then followed by the \emph{Resampling} step with a gain of 11.9$\times$. 
Although the gain in the \emph{Resampling} step has less of an impact compared with 
the overwhelming gain in \emph{CDF} and \emph{Propagation}, 
it does not overshadow the fact that it gained 2.7$\times$ in single precision 
even if we ignore the time spent in sorting the uniform random variates and 
focus on the resampling procedure only. That implies that our parallel resampling on the GPU 
can beat the stratified resampling on the CPU since the stratified resampling is roughly 
equivalent to the resampling with sorted uniform variates without sorting 
in terms of computational complexity. 
As for the \emph{Other} step, \emph{CPU(s)} and \emph{GPU(sp)} is identical. 
This is because for both algorithms, all executions of this step are conducted only on the CPU. 
Thus, we observe no difference. 
Finally, we observe a good amount of reduction in initiating the particle learning algorithm
by our parallel algorithm; 
however, the time spent in initiation is quite trivial, 
in particular when the number of the sample period $T$ is large ($T=100$ in our experiment).

Although it is clear that our parallel algorithm is superior to 
the conventional sequential algorithm through every step, 
Table \ref{table.breakdown} indicates that there is one drawback of using the GPU; memory transfer. 
The \emph{Store} step measures the time it takes to transfer the generated particles 
from the GPU's device memory to the CPU's host memory. 
Table \ref{table.breakdown} shows that it takes up roughly 15-20\% of the execution time. 
Note that, for fairness of the experiment, the GPU returns all of the particles it generated 
to the CPU's host memory. If we were to return only the mean, the variance, and other statistics 
of the state variables and parameters, the time for the \emph{Store} step 
can be cut down significantly. 

\section{CONCLUSION}

In this study, we have developed a new parallel resampling algorithm for the first fully parallel algorithm to perform particle filtering and learning 
in a parallel computing environment.
Our new algorithm has several advantages. First, it enables us to keep all executions of the particle 
filtering (and learning) algorithm within the GPU so that data transfer between the GPU's device memory 
and the CPU's host memory is minimized. Second, unlike the stratified resampling or the systematic 
resampling, our parallel resampling algorithm based on the cut-point method can resample particles 
exactly from their CDF. Lastly, since our algorithm does not utilize any device specific 
functionalities and is algorithmically parallel, it is straightforward to apply our algorithm to a multiple GPU system or 
a large grid computing system.

Then we conducted a Monte Carlo experiment in order to compare our parallel algorithm with 
conventional sequential algorithms. In the experiment, our algorithm implemented on the GPU 
yields results far better than the conventional sequential algorithms on the CPU. 
Although we keep the SSM as simple as possible in the experiment, 
our parallel algorithm can also be applied to more complex models without 
any fundamental modifications to the programming code and this little investment will return 
a significant gain in execution time instantaneously.

Our fully parallelized particle filtering algorithm is beneficial for various applications 
that require estimating powerful but complex models in a very short span of time; 
ranging from motion tracking technology to high-frequency trading.
We even envision that one can perform real-time filtering of the state variables and 
the unknown parameters in a high-dimensional nonlinear non-Gaussian SSM on an affordable 
parallel computing system in a completely parallel manner.
That would pave the way for a new era of computationally intensive data analysis.

\bibliographystyle{aer}
\bibliography{reference}
\end{document}